\documentclass[aps,pra,amsmath,amssymb,twocolumn,reprint,showpacs,showkeys]{revtex4-1}
\usepackage{graphicx}
\usepackage{bm}
\usepackage{amsfonts}
\usepackage{xcolor,soul}
\usepackage{bbm}
\usepackage{amsthm}
\usepackage{bbm}
\usepackage{mathrsfs}  
\usepackage{hyperref}

	\newcommand{\ncd}{\newcommand}
	\ncd{\mrm}    {\mathrm}
	\ncd{\beq} {\begin{equation}}
	\ncd{\eeq} {\end{equation}}
	\ncd{\nn}{\nonumber}

	\def\d{{\rm d}}

	\def\iiota{\dot\iota}

\makeatletter
\def\@fnsymbol#1{\ensuremath{\ifcase#1\or \dagger\or *\or \ddagger\or
   \mathsection\or \mathparagraph\or \|\or **\or \dagger\dagger
   \or \ddagger\ddagger \else\@ctrerr\fi}}
    \makeatother
    
\makeatletter
\def\@pacs@name{MSC: }%
\makeatother

\begin{document}

\title{The geometry of induced currents in two dimensional media}

\author{Cesar S.~\surname{Lopez-Monsalvo}}
\email[]{cslm@azc.uam.mx}
\affiliation{Universidad Aut\'onoma Metropolitana Azcapotzalco, Av. San Pablo No. 420, Col. Nueva el Rosario, Alc. Azcapotzalco, C.P. 02128, Ciudad de M\'exico, Mexico}

\author{Servando \surname{Vargas-Serdio}}
\email[]{servandovs@azc.uam.mx (corresponding author)}
\affiliation{Universidad Aut\'onoma Metropolitana Azcapotzalco, Av. San Pablo No. 420, Col. Nueva el Rosario, Alc. Azcapotzalco, C.P. 02128, Ciudad de M\'exico, Mexico}

\author{Julian A.~\surname{Alzate-Cardenas}}
\email[]{jalzate@ictp.it}
\affiliation{The Abdus Salam International Centre for Theoretical Physics, Strada Costiera 11,
	Trieste 34151, Italy}

\author{Daniel \surname{Flores-Alfonso}}
\email[]{danflores@unap.cl}
\affiliation{Instituto de Ciencias Exactas y Naturales, Universidad Arturo Prat, Avenida Playa Brava 3256, 1111346, Iquique, Chile}
\affiliation{Facultad de Ciencias, Universidad Arturo Prat, Avenida Arturo Prat Chac\'on 2120, 1110939, Iquique, Chile}

\begin{abstract}

We present a framework that allows us to clearly identify the geometric features underlying the phenomenon of superconductivity in two dimensional materials. In particular, we show that any such medium whose response to an externally applied electromagnetic field is a geodesically flowing induced current, must be a superconductor. In this manner, we conclude that the underlying geometry of this type of media is that of a Lorentzian contact manifold. Moreover, we show that the macroscopic hallmark of their superconducting state is a purely topological condition equivalent to the geodesic nature of the induced current: the non-vanishing of its helicity.

\end{abstract}

\keywords{geodesic flow, contact geometry, 2D superconductivity, geometric electrodynamics}

\maketitle

\section{Introduction}

The physics and geometry of low dimensional superconducting media has rapidly become an outstanding edge in the general theory of superconductivity \cite{macdonald2019bilayer}.  Following the discovery of superconducting behaviour in twisted bilayered graphene \cite{cao2018unconventional, cao2018correlated}, there has been a growing interest in the electromagnetic properties  of 2D materials \cite{torma2022superconductivity, salamon2022flat, hofmann2023superconductivity}. Since superconductivity is an intrinsically quantum phenomenon, most of these efforts have been  concentrated in its associated microphysics \cite{simon2022role, mao2023diamagnetic}, leaving the macroscopic description limited to the fulfilment of London's equations. In this sense, a geometric theory  of the superconducting state --  from a non-quantum perspective -- is still lacking. Nevertheless, there has been a number of efforts in combining the geometric description of materials with the properties of various solutions of low dimensional gravity \cite{cvetivc2012graphene, gallerati2021negative}. These works, however, provide us with a promising arena to test effects stemming from quantum field theory in curved spacetimes which we will not pursue here. Instead,   in this work we present the \emph{macroscopic}  formulation of the intrinsic geometry of two dimensional superconducting materials, exhibiting the key geometric structures encoding superconductivity in such media  in a unified manner.

From a macroscopic point of view, the response of a given material to an external electromagnetic stimulus is prescribed through the \emph{constitutive relations}, expressing the \emph{induced} fields and currents in terms of the externally applied electromagnetic field. The simplest manner of describing a material medium is by considering the regime where its response is \emph{linear} and it is within that context where such linear constitutive relations are tantamount to the choice of a \emph{spacetime} metric tensor \cite{hehl2003foundations, sternberg2013curvature, lopez2020geometry, garcia2022light}.

That the response functions of a material medium can be expressed in terms of a \emph{Lorentzian} metric tensor was originally motivated by the fact that a class of spacetimes in General Relativity appear as effective optical media for the propagation of light \cite{gordon1923lichtfortpflanzung}. However, light propagation alone is not sufficient to determine in a unique manner a metric tensor and, hence, the corresponding analogue material. Thus, in order to establish a suitable correspondence between geometric objects and physical properties one also needs to consider the propagation of massive particles. In particular, the dynamics of charged particles when the medium is exposed to an electromagnetic field constrain further the geometric degrees of freedom.

It is worth noting that the \emph{extrinsic} curvature of the two-dimensional medium -- the one associated with `visible' deformations as seen from the surrounding space -- is not related to the intrinsic curvature -- the one associated with the material metric characterising the response of the medium \cite{KATANAEV1992}. The intrinsic curvature corresponds to the Riemann scalar of the material metric, emerging from the microscopic structure of the two-dimensional medium and from its response to external electromagnetic stimuli \cite{cvetivc2012graphene}. Thus, we focus on modelling 2D electromagnetic media using intrinsic geometry alone. 

Ever since the synthesis of graphene, theorists have been using the material as a laboratory to test relativistic ideas \cite{katsnelson2006chiral}. In this context, it becomes relevant to employ  differential geometry to describe 2D materials in order to reduce the gap between condensed matter and high energy physics. Such approaches have revealed a close relationship between graphene-like materials and (2+1)-dimensional black holes \cite{cvetivc2012graphene, gallerati2021negative}. What is more, the experimental issues arising in these descriptions of curved graphene have also been analysed and proposals exist for using techniques such as Scanning Tunnelling Microscopy to measure the analogs of phenomena like the Hawking-Unruh effect \cite{iorio2012hawkingunruh, iorio2014quantum}.

In this work we will assume that  the region occupied by the material is equipped with  a Lorentzian metric characterising in a macroscopic manner the  response functions of the material at the linear level, so that  the constitutive relations themselves are expressed via its corresponding Hodge dual operator (details of this well known formalism and can be found in standard references \cite{baldomir1996geometry, hehl2003foundations, sternberg2013curvature, zangwill2013modern, lopez2020geometry}, as well as in dedicated explorations for the two-dimensional case such as \cite{cvetivc2012graphene, cariglia2017curvature, gallerati2021negative}). Thus, we can ponder on the significance of the following questions: 1. \emph{Is there a type of media such that its response to an external electromagnetic field yields geodesic motion for the charge carriers?} and 2. \emph{What are the properties of such media?}

\section{Geodesic conditions for the charge carriers in the medium}

Let us consider a $(2 + 1)$-dimensional manifold $M$ endowed with a Lorentzian metric $g$ characterising an electromagnetic material filling a region $\Omega \subset M$ with boundary $\partial \Omega$ together with  an \emph{externally} applied electromagnetic field strength $F$. Here, externally applied means that $\Omega$ is immersed in the field $F$  controlled by means of external sources, i.e.  that are not themselves contained in the material region. In this sense, the medium is defined by the pair $(\Omega,g)$ whilst $F$ constitutes the net field strength produced by the external sources together with the response of the medium. For the present study we are not interested in the \emph{ambient} metric outside $\Omega$. Thus, to answer our questions, let us first obtain the dynamic conditions such that the charge carriers in $\Omega$ undergo geodesic motion in the presence of $F$.

Consider a vector field defined in the closure of $\Omega$, $\xi \in T\bar \Omega$, such that its integral curves describe the motion of the charge carriers in $\Omega$ and satisfies the timelike condition $g(\xi,\xi) = - c^2$ for a non-vanishing  function $c:\bar\Omega\rightarrow \mathbb{R}$ with units of velocity.  Here, $T\bar\Omega$ denotes the \emph{tangent bundle} of $\bar\Omega$. Thus, since the units associated with the metric tensor are those of  length squared, the vector field $\xi$ must have units of inverse time. The charges in the medium -- with mass $m$ and charge $q$ -- experience  the Lorentz force due to $F$ and other material forces regarded as a whole as `friction' due to the medium's electric resistance. We can express this as the geometric statement of Newton's dynamic equation  (cf. \cite{barros2005gauss, lopez2022exact}) 
	\begin{equation}
	\label{eq.lorentz}
		m \left(\nabla_\xi \xi\right)^\flat \overset{!}{=} - q \iiota_\xi F - R^\flat,
	\end{equation}
where we shall use $\overset{!}{=}$ to denote  an ``empirical'' equality. Here, for any vector field $X$, $X^\flat \equiv g^\flat(X) \equiv g(X,\cdot)$ is its \emph{metric dual} 1-form and the symbol $\iiota_{X}$ is used to denote the \emph{interior product} of a differential form with the vector field $X$.

Let us explore the contents of equation \eqref{eq.lorentz}. In the left hand side, $\nabla_\xi \xi$ defines the acceleration of the velocity vector field $\xi$ by means of the \emph{connection} $\nabla$. The connection defines the notion of parallel transport in the manifold. A vector field $\xi$ is said to be \emph{self-parallel} if it is \emph{parallely propagated} along itself, that is, if it satisfies the equation $\nabla_\xi \xi = 0$ \cite{nakahara2018geometry}. In the case $\nabla$ is torsion free and metric compatible (meaning $\nabla g = 0$) we call it  the Levi-Civita connection of $g$ and refer to the integral curves of the vector field $\xi$ as \emph{geodesics}. In these terms, \emph{inertial motion} -- the one with no acceleration -- corresponds to a geodesic flow. 

Similarly, in the right hand side of \eqref{eq.lorentz}, the first term corresponds to the geometric version of the \emph{Lorentz force} \cite{lopez2022exact}  whilst we represent with $R$ the additional forces that act in sum like a single one associated with friction or resistance. The geometric nature of the two terms is significantly different. On the one hand, dissipative forces act opposite to the direction of motion while, on the other, the electromagnetic force due to $F$ has in general non-vanishing components transverse to $\xi$. %
In this sense, they cannot cancel one another and, hence, the condition for geodesic motion requires that both terms vanish independently. However, note first that dissipative scenarios can still be contemplated from a geometric point of view, for example, through non-geodesic motion. Second, in the electromagnetic setting, the vanishing of the Lorentz force does not require the complete absence of an external field, but only the transversality of the electromagnetic 2-form to the velocity vector field, i.e.  $\iiota_\xi F = 0$; rendering our question the simplest version of the general problem of magnetic curves on Riemannian manifolds \cite{gibbons2009stationary}.

A non-dissipative velocity vector field $\xi$ satisfying this transversality condition with the electromagnetic 2-form is called a \emph{force-free field} \cite{ChandraKendll1957} in the sense that charges following its integral curves do not experience the Lorentz force and, hence, their motion corresponds to a special class of geodesics. 

In order to analyse the properties of this kind of field $\xi$, we use the compatibility condition between the Lie  and the covariant derivatives (cf. Chapter 4 in \cite{arnoldkhesin2021}),
	\begin{equation}
	\label{eq.compat}
	\left(\nabla_\xi \xi \right)^\flat = \pounds_\xi \xi^\flat - \frac{1}{2} \d g(\xi, \xi),
	\end{equation}
were $\pounds$ and $\d$ denote the Lie and exterior derivatives, respectively. 
Using Cartan's identity  -- $\pounds_\xi \xi^\flat = \iiota_\xi \d \xi^\flat + \d \iiota_\xi \xi^\flat$ --  the first term in the right hand side of \eqref{eq.compat} is
	$
	\pounds_\xi \xi^\flat 	= -c^2 \iiota_\xi \d \eta - \iiota_\xi \left(\d c^2\wedge \eta\right) - \d \left(c^2 \iiota_\xi \eta \right),
	$
where $\eta\in T^*\bar\Omega$ is the normalized metric dual 1-form to $\xi$ satisfying $\iiota_\xi \eta = 1$, meaning that $\xi^\flat=g^\flat(\xi) = - c^2 \eta$, where $\eta$ has units of ${\rm time}$. 

Our use in the dynamic equation \eqref{eq.lorentz} of the field $\xi$ for which $g(\xi,\xi) = -c^2$ means that we are taking the function $c$ to constitute the normalization the velocity vector field has when its integral curves are parameterized \emph{by arc length}. The function $c^{-2}$ is by definition tied to the electromagnetic properties of $\Omega$ via $g$ and can even be expected to equal the product between the medium's electric permittivity with its magnetic permeability \cite{lopez2020geometry}. Thus, we regard the case of a \emph{homogeneous} material to entail having a constant value for $c$. In such scenarios, this reduces the geodesic condition to
	\begin{equation}
	\left( \nabla_\xi \xi\right)^\flat = -c^2 \iiota_\xi \d \eta = 0.
	\end{equation}

Hence, in a homogeneous medium, the necessary and sufficient conditions for a timelike vector field $\xi$ to be geodesic are 
	\begin{align}
	\label{eq.geocond}
 	\iiota_\xi \eta = 1 \quad \text{and} \quad \iiota_\xi \d \eta = 0.
	\end{align}
 
In what follows, we shall refer to  $\xi$ as  the \emph{co-moving} observer with the induced current, as we explain below. 

\section{Geodesic induced current}

A material medium reacting to an externally applied electromagnetic field such that the induced current is geodesic can be described as follows. First, the induced current must flow along the integral curves of a field $\xi$ satisfying \eqref{eq.geocond}, that is
	\begin{equation}
	\label{eq.sharp}
	j_{\rm ind} = \lambda \xi  \quad \text{with} \quad \d \lambda =0.
	\end{equation}

Here the proportionality factor $\lambda$, which we assume to be constant for simplicity, must have units of charge per length squared. Notice that the units of $j_{\rm ind}$ are those of the current  density, i.e.  $[{\rm charge}/{\rm time}] \times [1/{\rm area}]$. Therefore,  $\lambda \equiv nq$ measures the number density $n$ of charge carriers moving along the geodesic flow in $\Omega$ as a  response to $F$, alone. Let us refer to $j_{\rm ind}$ as the \emph{geodesic} induced current. Here, we will assume that in  case the external field is absent, $\lambda$ completely vanishes. It may also be the case that there is no geodesic current at all even when the field is present, but we have excluded such case by assuming  a non dissipative medium. 

The corresponding metric dual 1-form to the  geodesic induced current  vector field is
	\begin{equation}
	\label{eq.jflat}
		j^\flat_{\rm ind} \equiv g^\flat(j_{\rm ind}) = -\lambda c^2 \eta.
	\end{equation}
	
Now, a charge-current  distribution is best understood as an $(n-1)$-form, in our case a 2-form $J$, which satisfies the \emph{conservation law}
	\begin{equation}
	\label{eq.cons1}
	\oint_{\partial \Omega} J \overset{!}{=} 0 \quad \text{with} \quad J = J_{\rm ext} + i J_{\rm ind},
	\end{equation}
where $i=\sqrt{-1}$. By virtue of Stokes' theorem, it follows the local conservation law $\d J = 0$. The units of such 2-form are those of charge \cite{hehl2003foundations}. Thus, equation \eqref{eq.cons1} is the mathematical expression of charge conservation. Note that $J$ is in general composed of two elements: an external one -- sourcing $F$  \emph{when} the material is absent --  and an induced one, corresponding to the response of the material  \emph{once} it is immersed in the field. Albeit there is no measurable physical distinction between  their constituent charge carriers, it is postulated that each one is conserved separately, that is, $\d J_{\rm ext} =0$ and $\d J_{\rm ind}= 0$, which follows from  our convenient splitting  of $J$ into  `real' and `imaginary' parts. Moreover, from the outset we are assuming that there are no external charges in the material region, thus $J_{\rm ext}\vert_\Omega = 0$. Nevertheless, in order to have a non-null external stimulus,  $J_{\rm ext}$ must not vanish outside  $\Omega$. 

The (2+1) decomposition of $J_{\rm ind}$ with respect to the \emph{co-moving} observer $\xi$ is given by \cite{hehl2003foundations}
	\begin{equation}
	\label{eq.2p1decomp}
	J_{\rm ind} = \rho_{\rm ind} - \eta \wedge \mathcal{J}_{\rm ind},
	\end{equation}
where $\rho_{\rm ind} \equiv \iiota_\xi \left(\eta \wedge J_{\rm ind} \right)$ and $\mathcal{J}_{\rm ind} \equiv - \iiota_\xi J_{\rm ind}$  are the charge and current densities spread across directions transverse to $\xi$, respectively.

Now we make a \emph{critical} assumption in our geometric construction. Let us consider that the \emph{totality} of the induced current in the medium is geodesically flowing and, therefore, we identify the 1-form $\star J_{\rm ind}$ with  $j^\flat_{\rm ind}$.  This is indeed a \emph{constitutive} relation expressing that, in addition to being geodesic, the induced current $j_{\rm ind}$ should be consistent with  the conservation law \eqref{eq.cons1}. Therefore, postulating
	\begin{equation}
	\label{eq.constitutive1}
	  j^\flat_{\rm ind} \overset{!}{=} - v \star J_{\rm ind}, 
	\end{equation}
guarantees for $j_{\rm ind}$ a continuity equation and renders $\xi$ into a divergenceless field.  This follows from the definition of the divergence operator (cf. \cite{arnoldkhesin2021}) acting on $j_{\rm ind}$, namely 
	$
	{\rm div}(j_{\rm ind})\mu = \d\left(\star j^\flat_{\rm ind} \right)
	$
which, by virtue of \eqref{eq.constitutive1} and the conservation law \eqref{eq.cons1} is identically zero. Indeed, $\d\left(\star j^\flat_{\rm ind} \right) = v \d J_{\rm ind} = 0$. Here, $\mu$ is the volume form of $\Omega$ and $v$ is a characteristic parameter of the material carrying units of velocity, which we also take to be constant whenever the material is homogeneous (as we corroborate later on by exhibiting its link to the value of $c$). Note that the relation between the conserved objects $J_{\rm ind}$ and $j_{\rm ind}$ depends on the medium properties through the Hodge star operator associated  to the metric $g$.

We shall refer to \eqref{eq.constitutive1}  as the \emph{charge-current constitutive relation}. Such identification immediately yields $J_{\rm ind}$ transverse to $\xi$. One can see this by using [cf. equation \eqref{eq.2p1decomp}] 
$\mathcal{J}_{\rm ind} =  \star [(\star J_{\rm ind}) \wedge \xi^{\flat}] = 0$, so that $J_{\rm ind} = \rho_{\rm ind}$. In this sense, the co-moving observer does not `see' any current distribution at all. Additionally, it follows 
 that
	$
	v \star J_{\rm ind} =  \lambda c^2 \eta.
	$
Therefore, the induced current 2-form can be expressed in terms of the Hodge dual of $\eta$, i.e.
	\begin{equation}
	\label{eq.jind1}
	J_{\rm ind} = - \frac{\lambda c^2}{v} \star \eta,
	\end{equation}
in accordance with $\star\eta$ being also transverse to $\xi$.

Now, note that in a three dimensional manifold  the transverse 2-forms to $\xi$ can be generated by a single non-vanishing 2-form  which, in light of \eqref{eq.jind1}, we identify as $\star \eta$. This in particular applies to $\d\eta.$

Alternatively, notice that the geodesic condition  $\iiota_\xi (\star\star\d \eta) = 0$ [cf. equation \eqref{eq.geocond}] is equivalent to $\star[(\star\d\eta)\wedge\xi^{\flat}]=0$ which in turn implies $(\star\d \eta) \wedge\eta = 0$. Therefore,
	\begin{equation}
	\label{eq.assoc}
	\d \eta =  \beta \star \eta \quad \text{with} \quad \d\beta = 0,
	\end{equation}	
where the units of $\beta$ should be those of \emph{inverse length}. Also, for the sake of simplicity we assume the parameter $\beta$ to be constant. In addition, notice that equation \eqref{eq.assoc} is only valid for 3 dimensional manifolds. Furthermore, from  $\eta \wedge \star \eta = -c^{-2} \mu$, it immediately follows that
	\begin{equation}
	\label{eq.nonint}
	\eta\wedge \d \eta = - \frac{\beta}{c^2} \mu.
	\end{equation}

A geodesically flowing induced current in a $(2+1)$-dimensional  homogeneous  medium leaves us with two scenarios. On the one hand, the case $\beta\neq 0$ indicates that the 1-form $\eta$ is \emph{not} closed and, from \eqref{eq.nonint}, it must be \emph{non-integrable}. On the other hand, $\beta=0$ makes $\eta$  an harmonic form. Let us concentrate on the former scenario.

In the case $\beta\neq 0$, the charge carriers moving along the geodesics of $\xi$ see a non-integrable charge distribution. That is, the induced charge in the medium  is distributed in a collection of planes transverse to the geodesic motion, twisting in such a way that  no global surface of constant charge  is observed by $\xi$. Since the manifold is $(2+1)$-dimensional, it also means that $\eta$ is a \emph{contact form} while the geodesic conditions, equations \eqref{eq.geocond}, reveal that $\xi$ is the corresponding Reeb vector field (see \cite{geiges2008introduction} for an introduction to contact geometry and topology). Furthermore, the geodesic condition \eqref{eq.assoc} (typically referred to as a Beltrami or ``force-free'' condition \cite{Kholodenk2013applicContctGeom}) is fulfilled whenever the metric is associated and compatible with the contact structure \cite{calvaruso2011contact}. This fact was already observed in  \cite{flores2021contact}.  Therefore, a homogeneous 2D material such that its  induced current  follows the geodesic flow of a timelike vector field is modelled by a Lorentzian $(2+1)$-metric contact manifold. These manifolds can be interpreted as gravitational vacuum solutions where the contact structure describes the spacetime's vorticity 
\cite{calvaruso2011contact, flores2021contact}.

Combining \eqref{eq.jind1}  with the geodesic condition  \eqref{eq.assoc}  immediately yields
	\begin{equation}
	\label{mastereq:J}
		\d \star J_{\rm ind} = -\beta J_{\rm ind},
	\end{equation}
leading us to the condition 
	\begin{equation}
	\Delta J_{\rm ind} = -\beta^2 J_{\rm ind},
	\end{equation}
where $\Delta$ is the Laplace-de Rham operator associated with  $g$. Therefore, the geodesic flow of the induced current, together with its corresponding conservation law and the charge-current constitutive relation, yields a Lorentzian Helmholtz relation for $J_{\rm ind}$ which, in the stationary case, encompasses the relevant scenarios for which $j_{\rm ind}$ is tangent to $\partial\Omega$.

\section{Field equations in the medium}

Now we have to link $j_{\rm ind}$ with its originating field $F$. We concentrate on the case of materials which do not remain magnetized nor polarized when the field is turned off.  That is, we are demanding that whenever $F$ vanishes the  induced current $j_{\rm ind}$ does as well. 

The field strength 2-form $F$ obeys the conservation law
	\begin{equation}
	\label{eq.const2}
	\oint_{\partial \Sigma} F \overset{!}{=} 0,
	\end{equation}
where $\Sigma$ is any three dimensional sub-region of $M$ with boundary $\partial \Sigma$. Once more, by means of Stokes' theorem, this  expression corresponds to the local form of the homogeneous Maxwell equations, $\d F = 0$.  These equations imply the existence of a \emph{local} gauge potential 1-form $A$ such that $F=\d A$. Here, the units of $F$ are  those of action per charge.

The inhomogeneous Maxwell equations  arise from the conservation law \eqref{eq.cons1}. That is, there exists a local 1-form $G$  with units of charge -- the excitation field -- such that $J=\d G$. Such field is normally used to establish the connection between $F$ and $J$ via the  \emph{field constitutive relation}
	\begin{equation}
	\label{rel:constHodge}
	G \overset{!}{=} \frac{1}{Z} \star F,
	\end{equation}
where $Z\in \mathbb{C}$ is a complex scalar which -- in the (2+1)-dimensional case -- has units of  impedance per unit length. Let us stress that such units depend on the dimension of $M$ and, albeit the construction is generic, these must be adapted accordingly for the higher dimensional counterpart. Thus, the inhomogeneous Maxwell's equations are split into a \emph{complex}  relation between the external and induced currents with the field strength; the real and imaginary parts corresponding to the dielectric and conducting responses, respectively. In particular, in the material region $\Omega$ where we assumed $J_{\rm ext}\vert_\Omega = 0$, $Z$  becomes purely imaginary and we have 
	\begin{equation}
	\label{eq.indmax}
	J_{\rm ind} = \frac{1}{\zeta} \d \star F,
	\end{equation}
where $1/\zeta \equiv \Im(1/Z)$  is also a constant.

Proceeding as before, the (2+1)-decomposition of  $F$ in terms of the observer $\xi$ yields [cf. equation \eqref{eq.2p1decomp}]
	\begin{equation}
	F = B_\xi-\eta \wedge E_\xi,
	\end{equation}
where $B_\xi \equiv \iiota_\xi \left(\eta \wedge F \right)$ and $E_\xi \equiv -\iiota_\xi F$   are the `magnetic' and `electric' fields measured by the co-moving observer, respectively. We can immediately see that $B_\xi$ is a 2-form transverse to $\xi$ and, as we previously noted, it ought to be parallel to $\star \eta$, meaning $B_\xi = bc \star \eta$, where $b$ is a \emph{scalar} with units of magnetic flux, i.e. action per charge per length squared.  Additionally,  from equation \eqref{eq.lorentz} together with the geodesic condition \eqref{eq.geocond}, we have $E_\xi = 0$. Hence,  $F$ is purely transverse, that is,  it is written as
	\begin{equation}
	\label{eq.fstareta}
	F =  bc \star \eta.
	\end{equation}
 
 This geometric identification allows us to  see that $F$ and $J_{\rm ind}$ are parallel. Thus, combining equations  \eqref{eq.jind1} and \eqref{eq.fstareta}, we obtain  the linear relation
	\begin{equation}
	\label{eq.ohmhall}
	J_{\rm ind} = - \frac{\lambda c}{b v} F.
	\end{equation}

This can also be written  from the point of view of the co-moving observer as  $\rho_{\rm ind} = - \lambda c \left(bv \right)^{-1} B_\xi$, akin to the  Hall effect (cf. Section B.4.5 of \cite{hehl2003foundations} and Chapter 3 of \cite{simon2013oxford}). Also, note that \eqref{eq.ohmhall} resembles a geometric statement of Ohm's law, where the induced current is proportional to the field. 

It remains to establish the functional nature of $b =- c^{-1} \iiota_\xi \star F$. To this end, let us take the exterior derivative of \eqref{eq.ohmhall}. Since both, the induced current and the field strength are conserved quantities, it follows that  $\d F = -  v (\lambda c )^{-1} \d b \wedge J_{\rm ind}=0$ and, hence, $\iiota_\xi \left(\d b \wedge J_{\rm ind}\right)=0$. Recalling that $J_{\rm ind}$ is transverse to $\xi$, we obtain  that $\iiota_\xi \d b =0$. Now, using Maxwell's equation \eqref{eq.indmax} together with 
 \eqref{eq.assoc}, it follows that  $\d \star F = -c(\d b \wedge \eta + \beta b \star \eta) = \zeta J_{\rm ind}$. Therefore, we get an expression relating all the factors our geometric construction has required, i.e. 
	\begin{equation}
	 \label{rel:betandb}
	  v\beta b =c  \zeta\lambda 
	\end{equation}
along with the condition $\d b \wedge \eta = 0$. We immediately see that $\d b = k\eta$, with $k=\iiota_\xi \d b = 0$; rendering  $b$ a non-vanishing \emph{constant} determining the intensity of $F$. Moreover, equation \eqref{rel:betandb} is fully consistent with our initial assumption for $\lambda$,   where we posed that it should vanish whenever the external field is absent.

Notice that, albeit in $\Omega$ there are no external sources, the magnitude of $F$ arises due to both, $J_{\rm ext}$ and $J_{\rm ind}$. In this sense $b$ corresponds  to  the magnitude of the combined external stimulus $b_{\rm ext}$ and its induced response $b_{\rm ind}$. That is,  $b =   b_{\rm ext} +  b_{\rm ind}$, where  $b_{\rm ind} = 0$ if $b_{\rm ext} = 0$.  Notice that the constancy of $b$ only implies that $\d b_{\rm ind} = -\d \left(b_{\rm ext} \right)$.  In this manner, equation \eqref{rel:betandb} can be written as
	\begin{equation}
	\frac{1}{\beta} = \frac{v}{\zeta \lambda c} \left(b_{\rm ext} + b_{\rm ind} \right),
	\end{equation}
 encoding various of the observed possible functional dependencies  between the constitutive parameters and the applied field \cite{Khasanov2014lambdaofbprb, Kadono2004lambdaofbJphysConds, Wang2022supercondgrapheNANO, Wu2019diamagnet2DPRB}.  

 The geometric nature of $F$, expressed solely in terms of $\star\eta$ [cf. equation \eqref{eq.fstareta}], allows us to replicate the argument used for  $J_{\rm ind}$ to obtain \eqref{mastereq:J}. In this case we have 
	\begin{equation}
	\label{eq.diamag}
	\d \star F = -\beta F,
	\end{equation}
leading once more to the Lorentzian Helmholtz equation 
	\begin{equation}
	\label{MeissnerF}
	\Delta F = -\beta^2 F.
	\end{equation}

Equation \eqref{MeissnerF} indicates that  a medium in which the induced current undergoes geodesic flow must exhibit perfect diamagnetism. In particular, from the co-moving observer point of view we have $\Delta B_\xi = -\beta^2 B_\xi$.  This is in complete agreement with the experimental evidence supporting the emergence of the Meissner effect in 2D superconductors  \cite{Wang2022supercondgrapheNANO, Yao2019diamagnet2DPRL, Wu2019diamagnet2DPRB, He2016diamagnetmeasur2D}.  This also reveals the identity  of  $\beta$ [introduced in  the geometric condition \eqref{eq.assoc}] as the inverse of the \emph{penetration depth}. Moreover, because we have not made any assumption about whether $F$ is observed to be ``purely magnetic'' or not from any particular `laboratory' frame, the scenario thus described by our equation \eqref{MeissnerF} constitutes a fully covariant description of the entire electromagnetic field. This is in accordance with other treatments on the matter like those of \cite{Londons1935supracond, Salasnich2024}. Equation \eqref{MeissnerF} therefore means that our construction also contemplates for the electric part of $F$ a penetration depth given by $\beta^{-1}$. 

Equations \eqref{eq.ohmhall} and \eqref{eq.diamag}, together with the charge current constitutive relation \eqref{eq.constitutive1} also lead us  directly to
	\begin{equation}
	\label{eq.london1}
	\d j^{\flat}_{\rm ind} = -\frac{v\beta^2}{\zeta} F.
	\end{equation}

This is no other than the \emph{gauge independent} London's equation (cf. Chapter 12 in \cite{sternberg2013curvature}). Moreover,  we can immediately 
obtain the \emph{gauge dependent} relation
	\begin{equation}
	\label{eq.london2}
	j^\flat_{\rm ind} = - \frac{v\beta^2}{\zeta} A + \psi \quad \text{with} \quad \d \psi = 0,
	\end{equation}
where $\psi$ is a closed gauge 1-form.
Here,  one can directly identify  the \emph{superfluid weight}  
 as $D_s \equiv \left(v \beta^2/\zeta \right)$ whose presence alongside the condition of zero resistance -- as argued  in \cite{liang2017band,torma2022superconductivity} -- is the defining property of superconductors.
 
 Recall that in general the superfluid weight constitutes a matrix, or rather, a linear operator. In equation \eqref{eq.london2} it appears as a scalar, thus manifesting that the medium is isotropic, in addition to homogeneous. This isotropy is a consequence of limiting ourselves to the case where $\lambda$ and $\beta$ are constant.
Furthermore, the general scenario of a non-homogeneous and non-isotropic 2D material can simply be addressed by relaxing each of our uniformity constraints.
 
\section{Helicity and the electromagnetic action}

Thus far we have neglected the possibility of $\beta$ being zero, which means  having $\d \eta =0$.  Using again Maxwell's equations under the assumption of a \emph{non-vanishing}  geodesic induced current,  we have $\zeta J_{\rm ind} = \d \star F  =  - c \d b \wedge \eta$.  As pointed out already,  
 the  constitutive relation \eqref{eq.constitutive1} is the one responsible for the transversality of $J_{\rm ind}$ with respect to $\xi$.  Thus,  we get $\d b \wedge \eta = 0$  and $J_{\rm ind}= 0$. Since this is a contradiction to our assumption of a non-vanishing  $j_{\rm ind}$, we see that the  case $\beta = 0$ is inconsistent with our electrodynamic construction -- as indicated as well by \eqref{rel:betandb}.

Notice that we have been able to identify each constant appearing in our construction except for $v$. Albeit one can use \eqref{rel:betandb} to obtain its value in terms of all the others, there is still one more thing we can do to reveal its identity and close our discourse.  

Let us consider the action of the electromagnetic field in the material region 
	\begin{equation}
	\label{ea.action1}
	S =- \frac{1}{2} \int_{\Omega} F \wedge G.
	\end{equation}

This can be written in terms of the squared \emph{norm} of $\eta$ or, alternatively, of $\xi$. To see this, one only needs to substitute in \eqref{ea.action1} the field constitutive relation \eqref{rel:constHodge} in combination with the linear relation \eqref{eq.ohmhall}, followed by the current constitutive relation \eqref{eq.constitutive1} so that
	\begin{equation}
	\label{eq.action2}
	S =  \frac{1}{2}\int_{\Omega} \frac{\zeta \lambda^2 c^4}{v^2 \beta ^2}\  \eta \wedge \star \eta = \int_{\Omega} \frac{1}{2 v} \varrho_s g(\xi,\xi) \mu,
	\end{equation}
where 
	$
	\varrho_{\rm s} \equiv \lambda^2 \zeta/v \beta^2 \equiv n m
	$
is the mass density of the \emph{superconducting} charge carriers. Indeed, $\varrho_{\rm s}$ has  units of mass per unit area while $m$ is the mass appearing in Newton's dynamic law [cf. equation \eqref{eq.lorentz}]. Recalling that $\mu$ carries a geometric factor of $c$ in any orthonormal frame, the invariant nature of the action \eqref{eq.action2}  reveals that $v$ is  precisely  $c$. Notice that this reproduces the usual expression for the penetration depth \cite{Salasnich2024}
	\begin{equation}
		\frac{1}{\beta^{2}} = \frac{c}{\zeta}\frac{m}{ n q^2 }.
	\end{equation}

\section{Final remarks}

Let us close our work by observing that the geodesic condition \eqref{eq.assoc} extremises the action \eqref{eq.action2} under volume preserving diffeomorphisms -- that is, under variations that  ensure the conservation of charge -- whenever $j_{\rm ind}$ is tangent to $\partial \Omega$, as in the Meissner state.  This fact exhibits that both, equations \eqref{mastereq:J} and \eqref{eq.diamag}, are extrema of the  electromagnetic action \eqref{ea.action1}. Furthermore, the integrand of the right hand side of \eqref{eq.action2} is simply the Lagrangian density of a perfect fluid, whose corresponding Euler-Lagrange equations -- subject to  the conservation of the induced current  -- encompass the divergenceless  material geodesics we started from  \cite{de1992relativity, arnoldkhesin2021, lopez2011thermal}.

As a final remark we see that the \emph{on-shell} electromagnetic field action of these 2D superconductors is proportional to the helicity of the induced current
	\begin{equation}
	\label{eq.helicity}
	\mathcal{H}(j_{\rm ind}) = \int_{\Omega} \left(\lambda^2 c^4 \right)\ \eta \wedge \d \eta = -\beta \lambda^2 c^2 {\rm vol}_\Omega.
	\end{equation}

This fact confers a topological nature to the action since the helicity does not depend on the metric (cf. \cite{arnold2014asymptotic} and Theorem 1.15 in \cite{arnoldkhesin2021}), i.e. only  the particular values of the constitutive constants distinguish one superconductor from another. Additionally, since all the dynamics -- field and current -- reduce to the geodesic condition \eqref{eq.assoc}, this suggests a ``field-current equivalence'' in which the current linkage [measured by  \eqref{eq.helicity}] together with the constitutive constants, completely characterize the superconducting state.


In sum, we conclude that the geometric hallmark of superconductivity in homogeneous and isotropic 2D materials is the  geodesic flow of a divergenceless induced current or, equivalently, the non-vanishing of its helicity.

\vskip0.5cm


{\it Acknowledgements.} We thank Alberto Rubio Ponce and the members of the Laboratorio de F\'isica Aplicada a la Existencia for fruitful discussions. CSLM would like to thank Centenario 107 for the support that resulted in the writing of this work.  SVS was funded through a Conahcyt post-doctoral fellowship Estancias Posdoctorales por M\'exico 2022(1). JAAC is thankful to Banco Santander for  economic support through an international studentship and to Universidad Aut\'onoma Metropolitana Azcapotzalco for his academic visit to Mexico that promoted the making of this work. DF is supported by Agencia Nacional de Investigaci\'on y Desarrollo under grant FONDECYT No. 3220083. 

{\it Author contributions.} {\it Lopez-Monsalvo}: Conceptualization, Methodology, Validation, Formal analysis, Resources, Writing - Original Draft, Supervision, Project administration. {\it Vargas-Serdio}: Methodology, Validation, Formal analysis, Investigation, Data Curation, Writing - Review \& Editing, Visualization. {\it Alzate-Cardenas}: Conceptualization, Methodology, Validation, Formal analysis, Investigation, Writing - Review \& Editing. {\it Flores-Alfonso}: Conceptualization, Methodology, Validation, Formal analysis, Writing - Review \& Editing, Supervision.

{\it Declarations of interest.} None.


\bibliographystyle{unsrt}
\bibliography{supercond}

\end{document}